# Vehicle to Vehicle (V2V) Communication Protocol: Components, Benefits, Challenges, Safety and Machine Learning Applications

Ramya Daddanala, Vekata Mannava
*School of Copmuter and Info. Sciences*
*University of the Cumberlands,*
KY, USA
rdaddanala2329@ucumberlands.edu,
vmannava27526@ucumberlands.edu

Lo'ai Tawlbeh, Mohammad Al-Ramahi
*Department of Computing and Cyber Security*
*Texas A&M University-San Antonio*
San Antonio, USA
Loai.Tawalbeh@tamusa.edu
Mohammad.Abdel@tamusa.edu

*Abstract*—**Vehicle to vehicle communication is a new technology that enables vehicles on roads to communicate with each other to reduce traffic, accidents and ensure the safety of people. The main objective of vehicle-to-vehicle communication protocol is to create an effective communication system for intelligent transport systems. The advancement in technology made vehicle industries to develop automatic vehicles that can share real-time information and protect each other from accidents. This research paper gives an explanation about the vehicle-to-vehicle communication process, benefits, and the challenges in enabling vehicle-to-vehicle communication as well as safety and machine learning applications.**

## I. INTRODUCTION

Vehicles are used mainly for transportation from one place to another. It saves time for people by transferring them to their destinations. Traffic on the roads is increasing in metropolitan areas. Vehicles are producing pollution as well as increasing accidents. Accidents happen every second on roads due to more traffic and due to violating traffic rules. To reduce accidents and traffic, new ideas are implemented by the government [1]. One of the trending technologies is to use Vehicle-to-every thing (V2X) communication.

Vehicle to everything (V2X) can be described as the mode of technology that facilitates the communication between a vehicle and its different moving parts of traffic operations that are around them such as different vehicles, the cyclists on the roads, the pedestrians who come around the vehicles, the buildings that are around the roads, the different road signs as well as the traffic lights on the roads. Through the vehicle to everything, the communication has been highly eased as it has been made to be faster, and even the communication, as well as other activities in the vehicles, has been automated [2].

Advancement in the technology, made vehicle-manufacturing industries to use machine learning algorithms and artificial intelligence and make automatic vehicles that can drive a car to any destination without any driver. V2V communication is sub category of V2X communication that allows advanced technology-based vehicles to share information about traffic, security, and safety using a specific protocols over a wireless network.

This technology can detect obstacles or any other medium through, combination of different types of alerts, say, audio, video or tactile. This way it warns the driver of any potential threats or even going further takes actions to prevent it. Usually these communications have a range of 300 meters and can detect anything in their way or anything received from another vehicle along as it is inside this range. Currently most of the vehicles with adaptive cruse and sensors and cameras or even radars, detect a threat to avoid a collision. But this system will soon be obsolete with growing demand for security in the vehicles which requires embedded hardware for cryptography [3]. V2V communication can help two vehicles to talk to each other and this way their chance for a crash or a traffic is very less (Figure 1) [4].

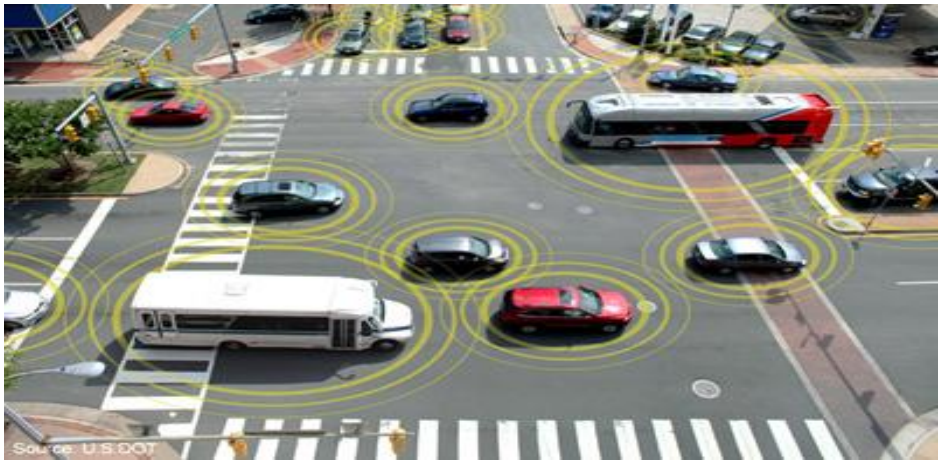

Fig. 1. What is Vehicle-to-Vehicle (V2V) communication [4]

All kinds of vehicles, including trucks, cars, motorcycles can use V2V communication, to take a step forward even bicycles can use V2V Communication. Also, as the vehicles speak to each other, the data being shared can also be controlled and any possible threats of data breaches are also minimal. V2V communication can increase the performance of a vehicle and also the driver security. As mentioned in [9], HTSA is dedicated to implementing this technology into today vehicles to counter the increasing accident rates and decreasing safety of the passengers.

II. HISTORY OF AN INTELLIGENT TRANSPORT SYSTEM

The United States has developed many plans for safe transportation. It initially created an interstate highway system to increase mobility in all cities and ensures safe transportation of goods. There were many issues at that time such as traffic congestion, highway fatalities, accidents, deaths, and impact on the environment due to high carbon emission. Later transportation professions and academia members of federal agencies discussed the transportation future and invented an intelligent vehicle highway system in 1990. It was renamed as an Intelligence transport system that can create efficiency in transportation when it is integrated with the latest technology in the future.

In future technologies like sensors, information systems, and mathematical methods can be used to increase the capacity of the Intelligence transport system to improve the transportation facility and reduce traffic congestion. In 1991, intermodal surface transportation efficiency act made the Intelligence transport system as its integral part and funded $660 to research on it more. The Intelligence transport system was renamed many times and its plans were changed many times, but the main objective of this system is to ensure mobility, security, and a healthy environment in the world.

Many Intelligence transport system programs were developed and implemented in the nation such as automated toll collections, traffic signals controls, and monitory systems to monitor the regions to address issues in real-time and so on. To improve safety on the roads and reduce crashes, collisions DSRC is used to connect the applications in the mobile environment. Later vehicle communication systems (Figure 2) [6] are invented to avoid road accidents by giving warning messages to drivers before they reach the dangerous path.

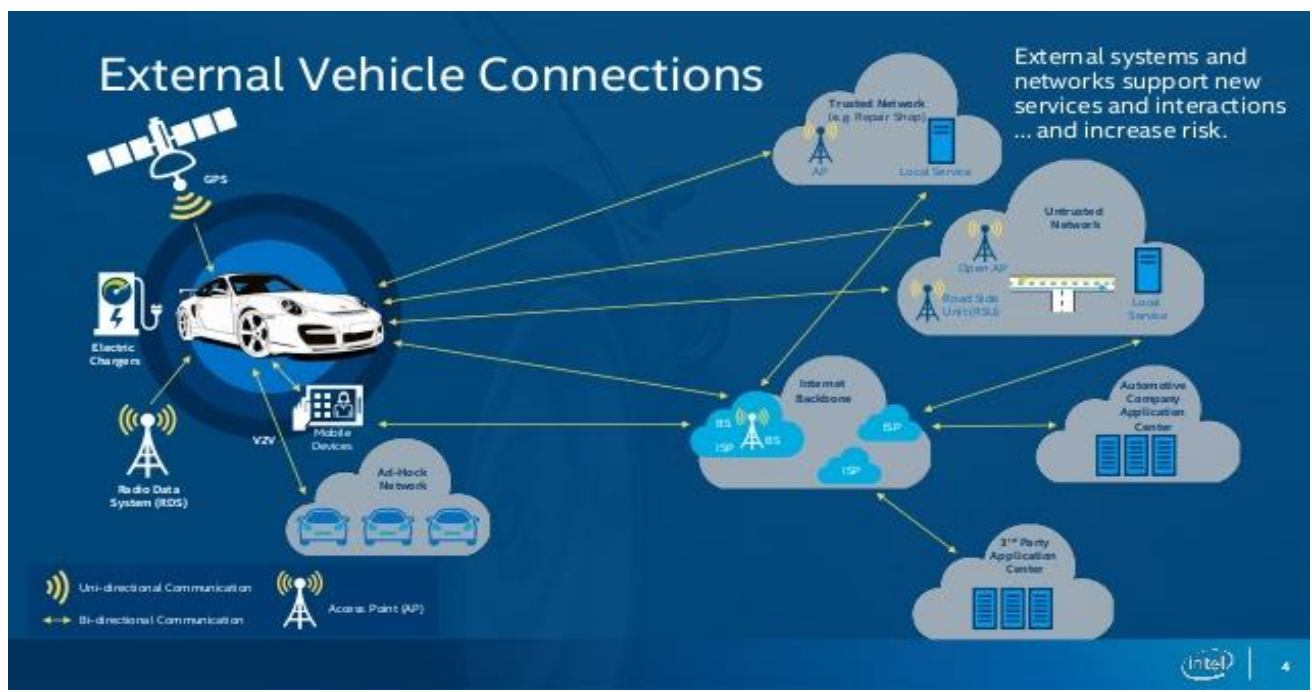

Fig. 2. Vehicle communication systems [6]

To reduce accidents and crashes on American roads, department of transportations of the federal government have issued a notice of proposed rulemaking that forces to enable vehicle-to-vehicle communication systems usages in the vehicle manufacturing process. This notice is used to ensure to send and receive messages from vehicles about the traffic and danger; it proposed a platform to communicate over a wireless network, to use the latest technologies in manufacturing vehicles, and to test them for ensuring safety on roads and avoid accidents. Vehicular Adhoc network has also gained more attention in recent years to avoid road accidents, traffic congestion, and to reduce fuel consumptions.

The intelligent transport system had developed this vehicle network to design safe and secure infrastructures for vehicles. The main objective of the vehicular Adhoc network is to provide safe transport management, traffic management, the security of passengers and to provide entertainment to the drivers in the journey. Vehicular Adhoc networks have different vehicle communication systems in it such as vehicle to vehicle communication, vehicle to infrastructure, vehicle to pedestrian, vehicle to the environment, vehicle to broadband cloud, and others that are used to provide valuable information to the driver and ensure safety on road. These communication systems use the latest technologies and depend on a wireless network to provide accurate information. The only thing to concern is that the wireless network must have high speed and bandwidth to reduce obstacles in communicating the message at the right time.

III. VEHICLE TO A VEHICLE COMMUNICATION SYSTEM

Vehicle to Vehicle (V2V) communication system research was started in 2003 under the vehicle infrastructure integration initiative program. Its origin was initiated in 1990 in automated highway systems that aimed to automate roadways completely and test the tracks. Vehicle to vehicle communication system uses the concept of internet of things (IoTs) and computing technologies [7].

The V2V communication helps vehicles to prevent accidents by allowing vehicles in transit to send and receive data about traffic using Adhoc mesh network. Either depending on the vehicle technology, the driver receives a warning message on road traffic or any accidental risks in front of the vehicle automatically takes action to change the root or to stop the vehicle in a safe location (Figure 3). According to researches and surveys, 60 percent of road accidents can be avoided in the world if the vehicle driver gets a warning message.

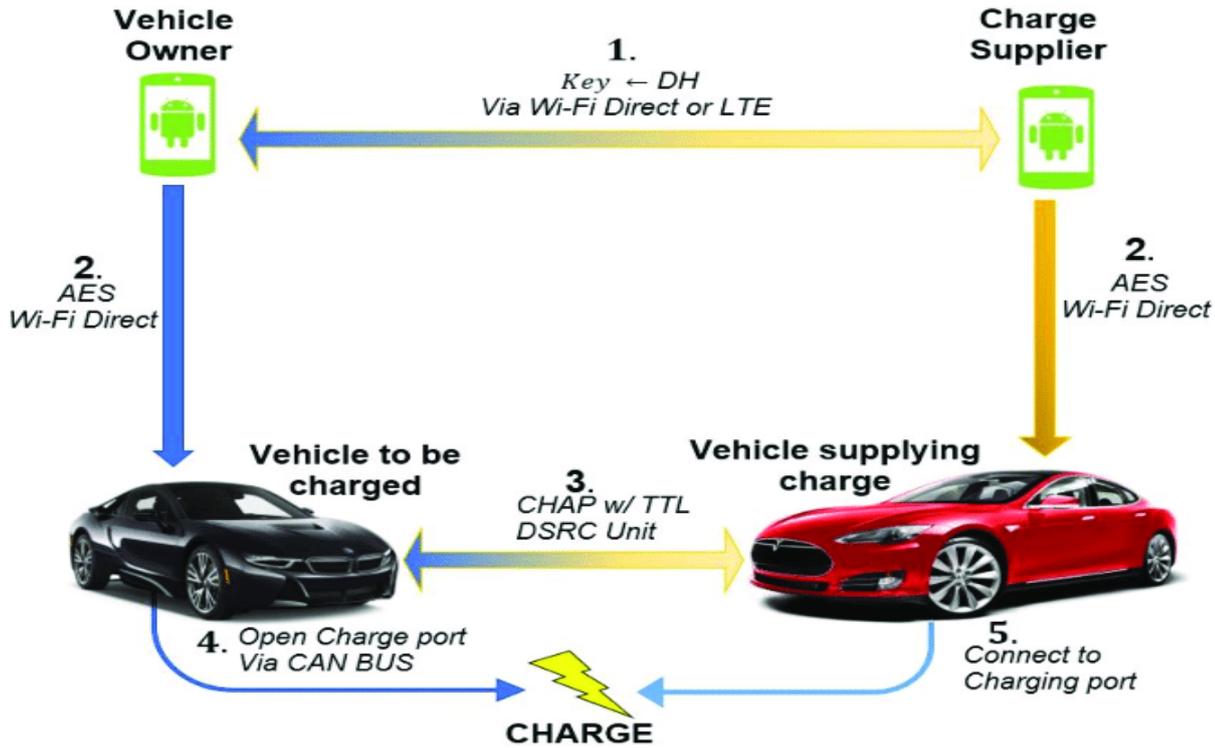

Fig. 3. Vehicle to vehicle communication system

This can be made possible using the vehicle-to-vehicle communication protocol and security. Emerging wireless technologies are used to reduce delays in sending emergency warnings. The dedicated short-range communication consortium helps to support public safety and private operations in a vehicle-to-vehicle communication environment. Vehicle-to-vehicle communication enables the cooperative collision warning among neighboring vehicles on roads and ensures the security and safety of peoples. However, vehicle-to-vehicle communication benefits humans in many ways but it can truly benefit if the wireless communication network has high speed and transformation of data to other vehicles. If there are any obstacles in communication paths and message is delivered in a wrong time then accidents may increases rather than decrease.

The vehicle-to-vehicle communication protocol uses broadcasts to send and receive warning messages. The messages are sent and received 10 times per second. If an appropriate software application is installed in the vehicles, then it can get messages from all its nearby vehicles and use it to avoid accidents and high traffic routes. The vehicle-to-vehicle communication messages have a range of 300 meters and detect danger using the weather condition, traffic, and terrain [11].

This technology also uses radars and cameras to detect and avoid collisions of vehicles. It not only helps drivers and automatic vehicles to survive from dangerous threats but also avoid creating crashes. Vehicle to vehicle communication protocols can be used by any vehicle like cars, buses, trucks to enhance their visibility. This communication protocol ensures the confidentiality of the driver's information while sending and receiving warning messages. The vehicle-to-vehicle communication protocol system increases the performance of the security system of vehicles and saves the lives of many. Every year millions of people die and get injures in road accidents, which can be avoided using this vehicle-to-vehicle communication protocol technology [11].

A. *components of vehicle to vehicle communication protocol [6]*

- Dedicated Short Range Communication (DSRC): it is a wireless communication channel designed for short-range use with a bandwidth of 75MHZ. It is used in vehicle-to-vehicle systems and vehicle to infrastructure systems to communicate information effectively.
- GPS receiver: It provides a real-time location of the vehicle that helps to navigate on the roads and find routes for different locations [12].

B. *Types of Vehicle to vehicle devices*

- OEM devices: It is an electronic device that is built in vehicles during the manufacturing process and the vehicle-to-vehicle communication system is connected to the devices and the data busses to give accurate messages to drivers. This device also processes the received messages, gives suggestions to drivers, and helps to avoid accidents and collisions.
- Aftermarket devices: Aftermarket devices are those devices that provide one or more functionalities in the vehicles in which it is installed during the manufacturing process. Some of the functionalities provided by it are comfort, safety, high performance, and convenience. The aftermarket devices with vehicle-to-vehicle communication systems communicate warning and notifications to the driver and help to avoid crashes. These devices are installed in the vehicle after the vehicle is manufactured completely. It is similar to a phone with an application. These devices can be used by any vehicle and even pedestrians can use this to avoid road accidents [6].
- Infrastructure based devices: These devices allow the vehicles to receive information from the infrastructure and update their data. It gives more accurate information about the traffic, and danger to the vehicle and reduces congestion. Vehicle to infrastructure is a complement of vehicle-to-vehicle communication. It gives different types of warnings to ensure the safety of the driver and the people inside the vehicle. Some of the warning given by this device is red light violation warning, speed zone warning, spot weather information warning, stop sign gap warning, curve speed warning, oversize vehicle warning, railroad crossing warning and so on. Figure 4 shows public key infrastructure for V2X [13].

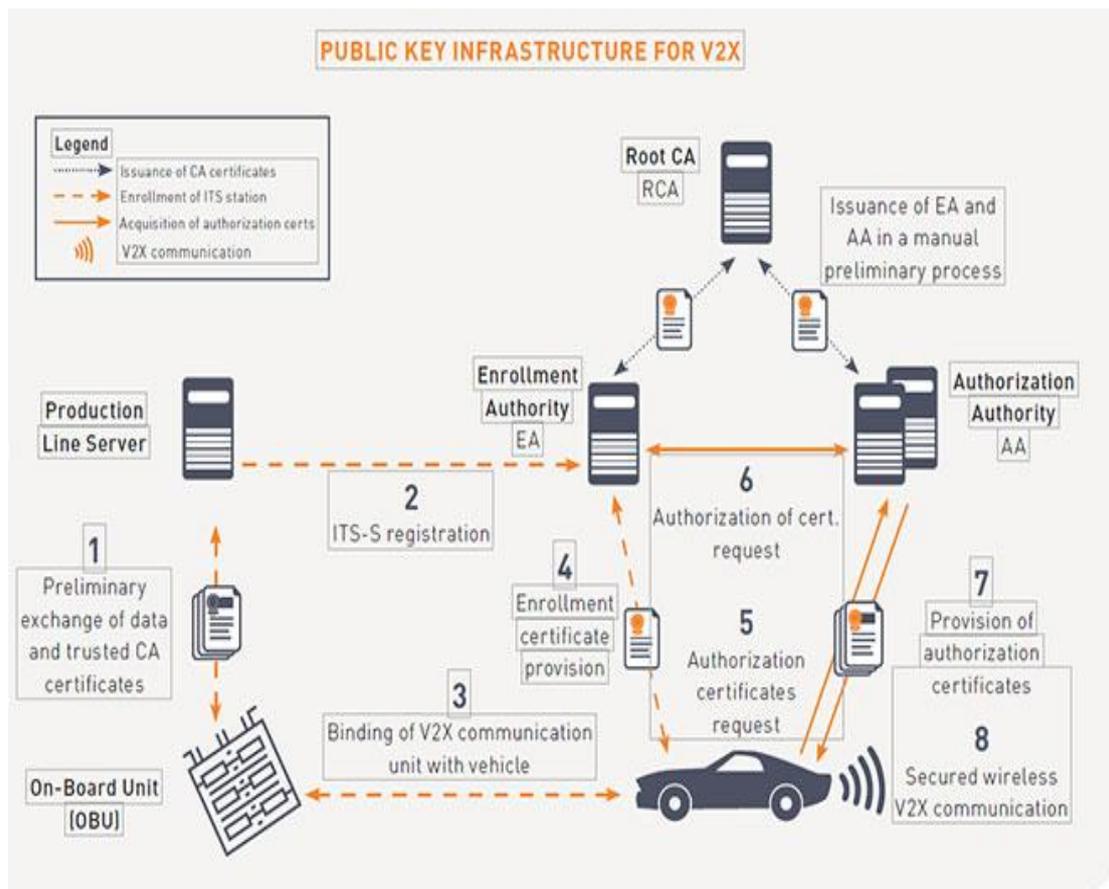

Fig. 4. Public key infrastructure for V2X [2]

IV. BENEFITS OF VEHICLE-TO-VEHICLE COMMUNICATION PROTOCOL

- Vehicle accidents kill millions of people every year. Accidents are increasing every day. A vehicle-to-vehicle communication protocol in vehicles prevents accidents by giving warning messages to drivers about the traffic and danger overhead.

- Traffic management is improved with this system and it helps to reduce congestion. Law enforcement officers use this vehicle-to-vehicle communication protocol to monitor the traffic using real-time data streams and manage traffic. They strictly impose rules and regulations and ensure the safety of people on the road. Drivers who use this protocol maintain distance from vehicles and reduce traffic jams. It also reduces the pollution emitted during the signals.

- It also helps to consume fuel effectively by the truck's platoons. Truck platoons are a group of trucks that follow each other. The leading truck will show the route that enables the other trucks to consume less fuel to reach the destination.

- The vehicle-to-vehicle communication protocol systems in vehicles will help to optimize the routes by giving exact location and less time-consuming route map to the drivers [13].

## V. LIMITATION OF VEHICLE-TO-VEHICLE COMMUNICATION PROTOCOL

Vehicle-to-vehicle communication protocol and security systems also have many limitations and impose challenges in implementing it in real-time. Some of the limitations are [9]:

- The frequency band allocated to this system will not support many vehicles at a time.

- It creates communication protocol issues, security issues in real-time that concern the public.

- Vehicles with Dedicated Short-Range Communication will create cyber-attacks on the system. It gives complete control to the attacker that create risks to live and identity threat. When the vehicle is attacked by the attacker, it will be used to create disasters, catastrophic, and used in terrorist attacks. To use this Vehicle-to-vehicle communication protocol and security systems in real-time, strong security measures must be used.

- The Vehicle-to-vehicle communication protocol and security systems store data about the vehicle location, vehicle details, driver details that can be misused by the person who has access to Automated License Plate Readers. The data can be used to theft identity and misuse the vehicle for wrong purposes.

- The Vehicle-to-vehicle communication protocol and security systems are not implemented yet and it is new to the world. It does not have any government regulations and rules. If the system miscommunicated information that leads to accidents, then the vehicle and the owner had to face financial losses.

- The Vehicle-to-vehicle communication protocol and security systems at present need human intervention to ensure safety. If the driver miscommunicated or gets distracted from his driving, then accidents may occur.

- The implementation of Vehicle-to-vehicle communication protocol and security systems in the vehicles depends on the model of the vehicle and the complexity of the system.

## VI. SAFETY APPLICATIONS OF VEHICLE-TO-VEHICLE COMMUNICATION SYSTEMS

There are various performance metric tools are present to measure the performance of the vehicle to vehicle communication systems that can be used to test the vehicle to vehicle system performance and use the results in creating federal motor vehicle safety standards. These metrics also stores information about the type of crash, the weather condition, the location, road condition, and others. This data is also used in researching about the crashes in the cities and helps to improve transportation facilities to reduce deaths. The safety applications are divided into four categories such as rear-end, opposite direction, junction cross, and lane change. The vehicle-to-vehicle safety applications are:

A. *Rear-end crash type warnings:*
- Forward collision warning: It will give warning to the driver about the rear-end collisions in the same lane and the direction to travel.

- Electronic emergency brake light: it warns the driver about the vehicle that is breaking the heard farther up ahead in the traffic and it needs to necessary to be in the same lane.

B. *Opposite direction crash type:*
- Do not pass warning: It gives warning to the dirtier if the passing zone is occupied by the other vehicle from the opposite direction in the same lane. It also advises the driver to move slowly to avoid a collision.

- Left turn assist: It gives warning to the driver when the other vehicle is turning left from the opposite direction to avoid a collision.

C. *Junction crossing crash type:*
- Intersection movement assists: it warns the driver when it is not safe to enter the intersection when there is a threat of collision with other vehicles.

D. *Lane change crash type:*
- Blindspot warning and lane change warning: it warns the driver when entering the blind zone about the vehicles coming from opposite and adjacent lanes to avoid crashes and accidents.

## VII. MACHINE LEARNING APPLICATIONS OF VEHICLE-TO-VEHICLE COMMUNICATION SYSTEMS

Machine learning provides efficient techniques and models to analyze large amount of data by discovering patterns and underlying structures in data. It represents an effective tool in different fields [e.g., 14, 15, 16]. In vehicular networks, machine learning is believed to be a promising solution to provide data-driven decisions in a wide variety of applications like *1) accurate and efficient* channel estimation in V2V wireless communications systems, 2) traffic flow prediction from real-time and historical traffic data collected by various on road and onboard sensors, 3) vehicle trajectory prediction, which is very critical for advanced driver assistance systems (ADAS) to perform many tasks such as road hazard warning and collision avoidance, 4) location prediction based scheduling and routing, 5) network congestion control, 6) load balancing and vertical handoff control, and 7) intrusion detection system to safeguard the information shared between vehicles [17].

## VIII. CONCLUSION

The Vehicle-to-vehicle communication protocol and security systems are in the developing stage. It is currently able to send and receive warning messages to and from drivers. In the future, this Vehicle-to-vehicle communication protocol and security systems can be used with automated vehicles that are manufactured using artificial intelligence and machine learning algorithms. In future automated vehicles become common and there will be no need for drivers to drive.

The distraction from driving and the security of data concerns will be eliminated. The Vehicle-to-vehicle communication protocol and security systems will increase driving efficiency, protect lives from accidents, and gives organizations to increase their productivity. It is also used to reduce the emission of carbon by reducing congestion in metropolitan cities. The paper has given a brief explanation about Vehicle-to-vehicle communication protocol and security systems, its benefits, limitations, and future aspects.

*Acknowledgment:* This research is supported by the Expanded Chancellor Research Initiative (CRI) grant awarded to Texas A&M University-San Antonio, TX, USA.